\documentstyle[prl,twocolumn,aps]{revtex}
\begin{document}
\draft

\title{Coherent Magnetotransport Through an Artificial Molecule}
\author{C.~A.~Stafford$^{1,*}$ and S.~Das Sarma$^2$}
\address{$\mbox{}^1$D\'{e}partement de Physique
Th\'{e}orique, Universit\'{e} de Gen\`{e}ve,
CH-1211 Gen\`{e}ve 4, Switzerland}
\address{$\mbox{}^2$Department of Physics,
University of Maryland, College Park, Maryland 20742}
\address{\rm (Received 26 April 1995)}
\address{\mbox{ }}

\address{\parbox{14cm}{\rm \mbox{ }\mbox{ }\mbox{ }
The conductance in an extended multiband
Hubbard model describing linear
arrays of up to ten quantum dots is calculated via a Lanczos technique.
A pronounced suppression of certain resonant conductance peaks
in an applied magnetic field due to a density-dependent
spin-polarization transition is predicted to be a clear signature of a
coherent ``molecular'' wavefunction in the array.
A many-body enhancement of localization
is predicted to give rise to a
{\em giant magnetoconductance} effect in systems with magnetic scattering.
}}
\address{\mbox{ }\mbox{ }}
\address{\parbox{14cm}{\rm PACS numbers:
73.20.Dx, 71.27.+a, 73.40.Gk, 78.66.-w}}

\maketitle

\makeatletter
\global\@specialpagefalse
\def\@oddhead{\underline{REV\TeX\mbox{ }3.0\hspace{11.7cm}
UGVA-DPT 1995 / 04-889}}
\let\@evenhead\@oddhead
\makeatother

\tighten
\narrowtext

Arrays of coupled quantum dots
\cite{dotsexp,waugh}
can be thought of as
systems of artificial atoms separated by tunable tunnel barriers.
Two complementary theoretical approaches have been useful in
describing such systems
in the limit \cite{classicaldots} where charging
effects are important but interdot tunneling is {\em incoherent} and
in the limit \cite{kirczenow} of coherent {\em ballistic} transport,
with charging effects neglected.
However, recent improvements in fabrication and experimental techniques
should make it possible to probe a third regime, where both interaction
{\em and} coherence effects play nontrivial roles.  In this regime,
the system of coupled quantum dots behaves like an artificial molecule,
and must be described by a
coherent many-body wavefunction \cite{us,klimeck}.
In this Letter, we describe some striking
characteristic signatures of such a coherent
molecular wavefunction in the low-temperature
magnetotransport through an array of quantum dots.
Our theoretical predictions should be experimentally testable in currently
available GaAs quantum dot systems.

An important consequence of coherent interdot tunneling is the formation of
interdot spin-spin correlations \cite{pwa}
analogous to those in a chemical bond
at an energy scale $\sim \, t^2/U$, where
$U=e^2/C_g$ is the charging energy of a quantum dot
and $t = (\hbar^2/2m^{\ast}) \int d^3 {\bf x} \,\Psi^{\ast}_i
({\bf x}) \nabla^2 \Psi_j ({\bf x})$ is the interdot hopping matrix element,
$\Psi_{i,j}$ being electronic orbitals on nearest-neighbor dots.
In an applied magnetic field, the Zeeman splitting drastically modifies the
many-body wavefunction of the array
when the Zeeman energy $g\mu_{B} B \sim t^2/U$, and we find that
certain resonant conductance peaks are suppressed (or, in some cases,
enhanced) by {\em several orders of magnitude} compared to their size at
$B=0$.
For quantum dots electrostatically defined in a 2D electron gas
\cite{dotsexp,waugh}, no significant modification of
the wavefunction of a single quantum dot would occur
for magnetic fields of this magnitude in the plane of the dots (the case we
consider),
so that the standard Coulomb blockade-based transport theory
\cite{classicaldots} which neglects
coherent interdot tunneling would predict {\em no} interesting magnetic field
dependence of the conductance.  Such spin-spin
correlations are intrinsic
many-body effects which are non-perturbative in the Coulomb interaction,
and can not be explained by
ballistic transport theories \cite{kirczenow} either.
We therefore believe that observation
of the predicted dramatic
magnetic field effect on low-temperature transport through coupled quantum
dots would represent the clearest possible signature of the formation
of an ``artificial molecule.''

The system we wish to model consists of a linear array of quantum dots
electrostatically defined \cite{dotsexp,waugh} in a 2D electron gas,
with a magnetic field {\em in the plane} of the dots.
We neglect intradot correlations (a reasonable approximation if the number of
electrons per dot is not too small \cite{us,onedot})
and focus instead on collective phenomena in the array.
The electron-electron interactions in the array are described
\cite{classicaldots,cap1} by a capacitance matrix $C_{ij}$:
we assume constant capacitances $C_g$ between each quantum dot and the
macroscopic
metallic gate which defines its confinement potential, and a capacitance
$C_i$, which is a function of gate voltage and may include important
quantum mechanical corrections \cite{cap1,cap2}, between nearest neighbor
quantum dots.
The electronic orbitals in the confining potential of an isolated dot
are taken to be nondegenerate with level spacing $\Delta$,
and the hopping matrix element $t_{n}$ between (nearly)
degenerate orbitals on nearest-neighbor dots is
assumed largest, all others being neglected.
The Hamiltonian is
\FL
\begin{eqnarray}
\mbox{ } & \!\!\!\! \hat{H} & = \sum_{j,n,\sigma}
\left(t_{n} \, \hat{c}_{j+1 n \sigma}^{\dagger} \hat{c}_{j n
\sigma} + \mbox{H.c.}\right)
\nonumber\\
& \!\!\!\! + &
\sum_{j,n,\sigma} \left(n\Delta + \sigma g \mu_{B} B/2\right)
\hat{c}_{j n \sigma}^{\dagger} \hat{c}_{j n \sigma}
+ \frac{e^2}{2} \sum_{i,j} C^{-1}_{ij} \hat{n}_{i} \hat{n}_{j},
\label{hubham}
\end{eqnarray}
where $\hat{c}^{\dagger}_{ j n \sigma}$ is the creation operator for an
electron of spin $\sigma$ in the $n$th orbital of the $j$th dot,
$\hat{n}_{j} \equiv \sum_{n,\sigma} \hat{c}_{j n \sigma}^{\dagger}
\hat{c}_{j n \sigma}$,
and the sums run from $n = 0$ to $M-1$ (the $M$ orbitals nearest
the Fermi energy $E_F$), $i,j = 1$ to $L$, and
$\sigma = \pm 1$.
In the strongly-correlated regime, the
interaction term in Eq.~(\ref{hubham}) cannot be treated perturbatively.
We therefore employ a Lanczos technique \cite{dagotto} to compute the
exact many body ground states of
(\ref{hubham}) for arrays of 5 to 12 quantum dots with 1 to 5
electronic orbitals per dot.

The array is coupled to noninteracting leads via a tunneling
Hamiltonian with matrix elements $t_{n}^{r,l} \propto t_{n}$
which couple
electrons in the $n$th orbital of the 1st ($L$th) dot to the right (left) lead.
The capacitance to the leads is neglected.
In the limit $\Delta E \gg k_B T \gg \hbar \Gamma$, where
$\Delta E$ is the energy level spacing in the array and
$\Gamma$ is the tunneling rate of electrons out of the array,
the linear response
conductance is determined by ground state to ground state
transitions, and is given by
\begin{equation}
G = e^2 \sum_{N} \frac{\Gamma_{N}^{r}
\Gamma_{N}^{l}}{\Gamma_{N}^{r} + \Gamma_{N}^{l}}
A_N (\mu),
\label{conductance}
\end{equation}
where $\Gamma_{N}^{r,l} = 2\pi\sum_{n,\sigma} |\langle 0_N|t_{n}^{r,l}
c_{1(L) n \sigma}^{\dagger}|0_{N-1} \rangle|^2 \rho^{r,l}(E_{N}^0 -
E_{N-1}^0)/\hbar$,
$\rho^{r,l}(\varepsilon)$ being the density of states in the leads.
For the case $B > 0$, the ground state is nondegenerate and
$A_N (\mu) = - f'(E_N^0 - E_{N-1}^0)$, while
for $B=0$, the ground state is spin-degenerate when $N$ is odd and
$A_N (\mu) = 2/[k_B T (3 + 2 e^{-x_N} + e^{x_N})]$, where $x_N = (-1)^N
(\mu -E_N^0 +E_{N-1}^0)/k_B T$.
Eq.~(\ref{conductance}) is derived by the method of
Refs.~\cite{klimeck,onedot}.

\begin{figure}
\vbox to 6.0cm {\vss\hbox to 8cm
 {\hss\
   {\includegraphics{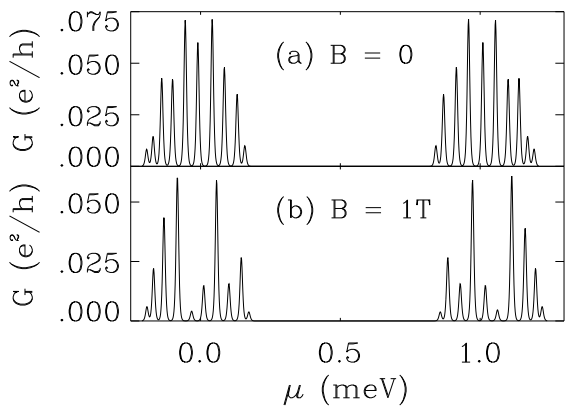} }
  \hss}
}
\caption{Conductance vs.\ chemical potential $\mu$ through a linear
array of 10 GaAs quantum dots with one spin-1/2 orbital per dot.
$e^2/C_g=1\mbox{meV}$, $C_i=0$,
$t=0.1\mbox{meV}$, and $T=35\mbox{mK}$.  Splitting of the two Coulomb
blockade peaks into minibands is driven by $t$.  The suppression of
the 5th peak in (b) is the result of a density-dependent
SPT.}
\label{fig1}
\end{figure}

Fig.\ \ref{fig1} shows the conductance through a linear array of 10 quantum
dots with $C_i =0$
as a function of the chemical potential $\mu$ in the leads, whose value
relative to the energy of the array is controlled by the gate voltages.
The two Coulomb blockade peaks in Fig.\ \ref{fig1}  are
split into multiplets of 10 by interdot tunneling, as discussed in
Refs.\ \onlinecite{us,klimeck}.  We refer to these multiplets as {\em
Hubbard minibands}.  The energy gap between
multiplets is caused by collective Coulomb blockade \cite{us},
and is analogous to the energy gap in a Mott
insulator \cite{andyandi}.  The heights of the resonant conductance peaks in
Fig.\ \ref{fig1}(a) can be understood as follows:
Since the barriers to the leads are assumed to be large, the single-particle
wavefunctions of
\noindent
the array are like those of a particle in a one-dimensional
box.  The lowest eigenstate has a maximum in the center
of the array and a long wavelength, hence a small amplitude on the end dots,
leading to a suppression of the 1st conductance peak.  Higher
energy single-particle states have shorter wavelengths, and hence larger
amplitudes on the end dots, leading to conductance peaks of increasing
height.  The suppression of the conductance peaks at the top of the 1st
miniband can be understood by an analogous argument in terms of many-body
eigenstates; the 10th electron which enters the array can be thought of as
filling a single hole in a Mott insulator, etc.

In Fig.\ \ref{fig1}(b), the spin-degeneracy of the quantum dot orbitals is
lifted by the Zeeman splitting.  There is a critical field $B_c$ above which
the system is spin-polarized; for an infinite 1D array with $C_i=0$ and $t
\ll U$ \cite{crit}
\begin{equation}
g\mu_{B} B_c \simeq \frac{4 t^2}{\pi U} (2 \pi n - \sin 2 \pi n),
\label{bcrit}
\end{equation}
where $n < 1$ is the filling factor of the lower Hubbard band.  (Recall that
we are here considering only the single spin-1/2 orbital nearest $E_F$
in each quantum dot---the magnetic field required to spin-polarize an entire
quantum dot is much larger \cite{klein}.)
Because $B_c$ is a function of $n$, one can pass through this spin-polarization
transition (SPT) by varying $n$ at fixed $B$.  In Fig.\ \ref{fig1}(b), this
transition occurs between the 4th and 5th electrons added
to the array, consistent with the prediction of Eq.\ (\ref{bcrit}).
The effect of this transition on the conductance spectrum is dramatic:
The first 4 electrons which enter the array have spin aligned with $B$ (up),
but the 5th electron enters with the opposite spin, and goes predominantly
into the lowest single-particle eigenstate for down-spin electrons, which
couples only weakly to the leads, leading to a suppression of the 5th
resonant conductance peak by over an order of magnitude.  It should be
emphasized that the heights of the conductance peaks change {\em
discontinuously} as a function of $B$ each time there is a spin-flip.

\begin{figure}
\vbox to 6.0cm {\vss\hbox to 8cm
 {\hss\
   {\includegraphics{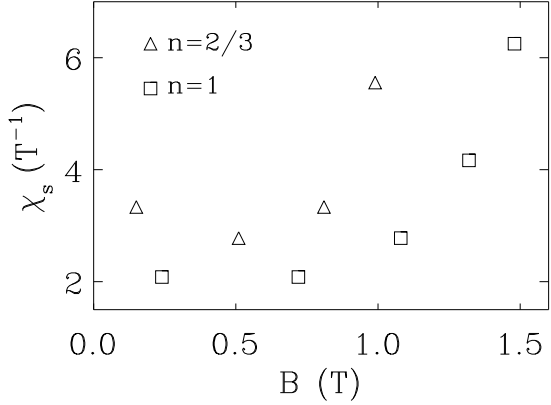} }
  \hss}
}
\caption{Spin susceptibility $\chi_s = \hbar^{-1} \Delta S/\Delta B$ at $T=0$
vs.\ magnetic field $B$ for linear arrays of GaAs
quantum dots with
$e^2/C_g=1\mbox{meV}$, $C_i/C_g=0.5$, and $t=.05\mbox{meV}$.
Squares:  10 electrons on 10 dots ($B_c \approx 1.5 \mbox{T}$);
triangles:  8 electrons on 12 dots  ($B_c \approx 1 \mbox{T}$).}
\label{chi_s}
\end{figure}

Splitting of the Coulomb blockade peaks due to interdot coupling
and suppression of the conductance peaks at the miniband edges
have recently been observed experimentally
by Waugh {\it et al.} \cite{waugh}.
However, it has been pointed out \cite{waugh} that both effects can also be
accounted for by a model \cite{classicaldots}
of capacitively coupled dots with completely
{\em incoherent} interdot tunneling.  It is therefore of
interest to consider the effects of interdot capacitive coupling in the
regime of {\em coherent} interdot transport.
A nonzero interdot capacitance $C_i$
introduces long-range electron-electron interactions in Eq.\ (\ref{hubham})
and decreases the intradot charging energy $U$.
Fig.\ \ref{chi_s} shows the spin susceptibility $\chi_s$
for $C_i/C_g=1/2$ in linear
arrays with 8 electrons on 12 dots and 10 electrons
on 10 dots.  The $n$-dependence of $B_c$ in Fig.\ \ref{chi_s}
is qualitatively similar to that in a system with intradot interactions only,
but the values of $B_c$ are roughly twice
those of a system with $C_i=0$.
Note the rapid growth of $\chi_s$ as $B \rightarrow B_c$.
In
\noindent
an infinite array, $\chi_s$ is expected to diverge as $B \rightarrow B_c$
because the system undergoes a second order quantum phase transition
\cite{crit}.  The SPT predicted to occur in an array of coupled
quantum dots is in contrast to that observed in a single quantum dot
\cite{klein}, where the critical point occurs for minimum total spin.

\begin{figure}
\vbox to 6.0cm {\vss\hbox to 8cm
 {\hss\
   {\includegraphics{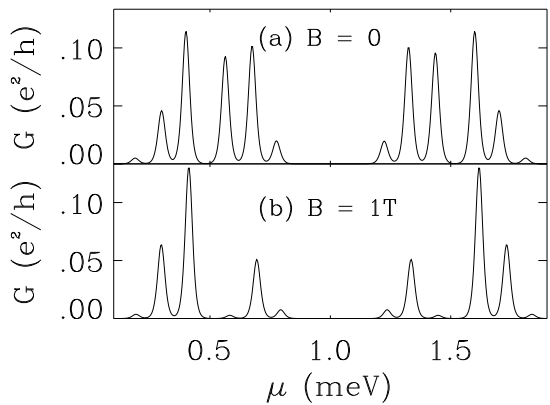} }
  \hss}
}
\caption{Conductance vs.\ chemical potential $\mu$ through a linear
array of 6 GaAs quantum dots with one spin-1/2 orbital per dot.
$e^2/C_g=1\mbox{meV}$, $C_i/C_g=0.5$,
$\bar{t}=.05\mbox{meV}$, $T=120\mbox{mK}$.  Disorder $\delta t/\bar{t}
\sim 1$ ($t_{i\uparrow}=t_{i\downarrow}$) is present in the hopping
matrix elements.  The splitting of the Coulomb blockade peaks into multiplets
is dominated by $C_i$; however, the effect of $B$ is similar to that in
Fig. 1.}
\label{fig2}
\end{figure}

Disorder introduces a length scale which cuts off the critical behavior
as $B \rightarrow B_c$.  However, as shown in Fig.\ \ref{fig2},
where disorder $\delta t \sim t$
has been included in the hopping matrix elements, the SPT
has a clear signature in the magnetotransport even in a
strongly disordered system.
In Fig.\ \ref{fig2}, the peak splitting due to
capacitive coupling is roughly ten times that due to interdot tunneling, so
that the peak positions are within $\sim 10\%$ of those predicted
\begin{figure}
\vbox to 6.0cm {\vss\hbox to 8cm
 {\hss\
   {\includegraphics{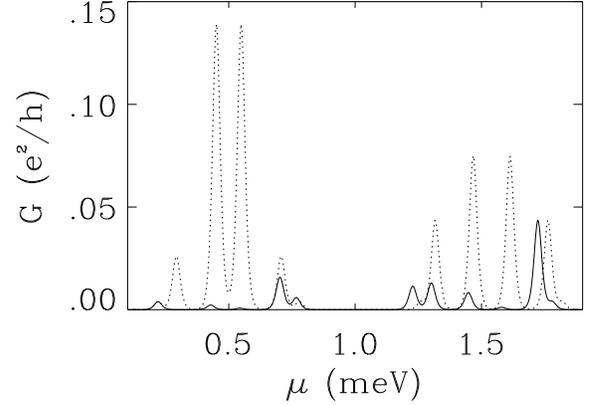} }
  \hss}
}
\caption{Conductance vs.\ chemical potential $\mu$ through a linear
array of 6 GaAs quantum dots with one spin-1/2 orbital per dot.
$e^2/C_g=1\mbox{meV}$, $C_i/C_g=0.5$,
$\bar{t}=.05\mbox{meV}$, $T=120\mbox{mK}$.  Spin-dependent
disorder $\delta t/\bar{t}
\sim 1$ ($t_{i\uparrow}\ne t_{i\downarrow}$) is
included in the hopping matrix elements.  Solid curve: $B=0$; dotted curve:
$B=1.3 \mbox{T}$.  At 1.3T, the 2nd conductance peak is enhanced by a factor
of 1600.}
\label{fig3}
\end{figure}
\noindent
by a classical charging model \cite{classicaldots}.  However, the dramatic
dependence of peak heights on magnetic field---the 4th conductance peak in
Fig.\ \ref{fig2}(b) is suppressed by a factor of 32 compared to its $B=0$
value due to the
density-dependent SPT described above---can not be accounted
for in a model which neglects coherent interdot
tunneling.  This effect should be observable provided $g\mu_{B} B_c >
\mbox{max}(k_B T,\hbar/\tau_i)$, where $\tau_i$ is the inelastic scattering
time.  We believe that this striking magnetotransport effect is the clearest
possible signature of a coherent molecular wavefunction in an array of
quantum dots.

Fig.\ \ref{fig3} shows the conductance spectrum for an array of 6 quantum
dots with the same parameters as in Fig.\ \ref{fig2}, but with
spin-dependent disorder in the hopping matrix elements, as could be
introduced by magnetic impurities.  Several conductance peaks at $B=0$
(solid curve) are strongly suppressed due to a many-body enhancement of
localization.  This effect arises because repulsive on-site interactions
enhance spin-density wave correlations, which are pinned by the
spin-dependent disorder \cite{tands}.  At $B=1.3 \mbox{T}$ (dotted curve)
the system is above $B_c$ and is spin-polarized, circumventing this effect.
The second conductance peak is enhanced by a factor of 1600 at 1.3T compared to
its size at $B=0$ (not visible on this scale).  This {\em giant
magnetoconductance} effect is a many-body effect
intrinsic to the regime of coherent interdot transport.

Another interesting phenomenon stemming from the competition between coherent
interdot tunneling and charging effects is the Mott-Hubbard metal-insulator
transition (MH-MIT), which occurs when collective Coulomb blockade (CCB)
\cite{us} is destroyed
due to strong interdot coupling.  For GaAs quantum dots larger than
about 100nm in diameter, we find that this transition is caused by the
divergence of the effective interdot
\begin{figure}
\vbox to 6.5cm {\vss\hbox to 8cm
 {\hss\
   {\includegraphics{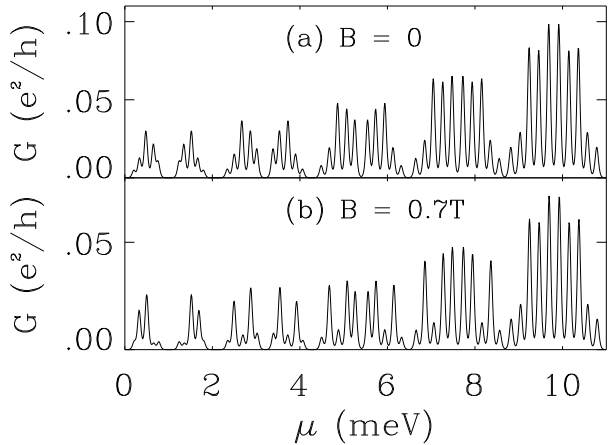} }
  \hss}
}
\caption{Conductance vs.\ chemical potential
through a linear array of 5 GaAs quantum
dots with 5 spin-1/2 orbitals per dot.
$e^2/C_g=1\,\mbox{meV}$, $\Delta = 0.2\,\mbox{meV}$, $T=.29\mbox{K}$,
$C_i^{(n)}/C_g=2^{n-1}$, and $t_{n} = 0.05\,\mbox{meV}
(1.05)^{n}$ ($n=0,\ldots,4$).
The energy gap between Hubbard minibands is not
resolved for $\mu > 9\,\mbox{meV}$ (breakdown of CCB).
Note the quenching of magnetoconductance effects in the ballistic regime.}
\label{fig.mit.longrange}
\end{figure}
\noindent
capacitance, similar to
the breakdown of Coulomb blockade in a single quantum dot \cite{cap2}.
Within the framework of the scaling theory of the MH-MIT
\cite{andyandi}, one
expects a crossover from CCB to ballistic transport in a finite array of
quantum dots when the correlation
length $\xi$ in the CCB phase significantly exceeds the linear dimension $L$
of the array.  Fig.\ \ref{fig.mit.longrange}
shows the conductance spectrum for 5 quantum dots
with 5 spin-1/2 orbitals per dot.
The divergence of the effective interdot capacitance as the interdot
barriers become transparent is simulated by setting
$C_i^{(n)}/C_g = 2^{n-1}$, $n=0,\ldots,4$.  In Fig.\ \ref{fig.mit.longrange},
minibands arising from each orbital are split symmetrically into multiplets
of 5 peaks by CCB, with the center to center spacing between
multiplets equal to $e^2/C_g$, while the energy gap
between minibands corresponds to the band gap $\sim \, \Delta$
enhanced by charging effects.
The CCB energy gap is evident in the first 3 minibands,
but is not resolvable for the
higher orbitals ($C_i/C_g \geq 4$),
although there is still a slight suppression of the
conductance peaks near the center of the 4th miniband.
Comparison of the compressibility of the system
to a universal scaling function for the MH-MIT calculated by the method of
Ref.\ \onlinecite{andyandi} indicates $\xi/L \sim 10^3$ for $C_i/C_g = 8$,
so that the transport in the 5th miniband is effectively ballistic.
The peak spacing within a miniband
saturates at $e^2/LC_g$ (plus quantum corrections $\sim t/L$)
in the ballistic phase
because the array behaves like one large capacitor, as observed experimentally
in Ref.\ \onlinecite{waugh}.
Fig.\ \ref{fig.mit.longrange}(b) shows the effects of a magnetic field on the
conductance spectrum:  a sequence of SPTs is
evident in the different minibands, with $B_c$ an
increasing function of $C_i/C_g$, leading to quenching of magnetoconductance
effects in the ballistic regime.

A finite-size scaling analysis of the compressibility indicates that the
MH-MIT probably occurs at $C_i/Cg =\infty$ in an infinite array of quantum
dots, when the interdot barriers become transparent to
one transmission mode \cite{smalldots}.

In conclusion, we predict that low-temperature
magnetotransport experiments on judiciously fabricated
quantum dot arrays would lead to the observation of a variety of
phenomena resulting from the interplay between coherent interdot
tunneling and charging effects:
SPT, MH-MIT, and giant magnetoconductance due to a many-body
enhancement of localization by magnetic scattering.

This work was supported by the U.S.\ Office of
Naval Research and the Swiss National Science Foundation.


\begin{references}
\bibitem{smalldots} In contrast, for coupled GaAs quantum dots smaller than
about 100nm in diameter, the breakdown of CCB is expected to occur \cite{us}
for $t \sim U$, when the transmission through the interdot barriers is less
than unity.
\end{references}
\end{document}